\documentclass[aps,prd,preprint,amsmath,amssymb,latexsym,array,enumerate,superscriptaddress]{revtex4}

\usepackage{amssymb}
\usepackage{amsmath}
\usepackage{epsfig}
\usepackage{hyperref}
\usepackage{breakurl}
\usepackage{xcolor}

\makeatletter
\def\simgt{\mathrel{\lower2.5pt\vbox{\lineskip=0pt\baselineskip=0pt
           \hbox{$>$}\hbox{$\sim$}}}}
\def\simlt{\mathrel{\lower2.5pt\vbox{\lineskip=0pt\baselineskip=0pt
           \hbox{$<$}\hbox{$\sim$}}}}
\makeatother

\newcommand{\be}{\begin{equation}}
\newcommand{\ee}{\end{equation}}
\newcommand{\bea}{\begin{eqnarray}}
\newcommand{\eea}{\end{eqnarray}}
\newcommand{\beq}{\begin{eqnarray}}
\newcommand{\eeq}{\end{eqnarray}}

\def\lsim{\mathrel{\rlap{\lower4pt\hbox{\hskip1pt$\sim$}}
     \raise1pt\hbox{$<$}}}         
\def\gsim{\mathrel{\rlap{\lower4pt\hbox{\hskip1pt$\sim$}}
     \raise1pt\hbox{$>$}}}         

\newcommand{\comment}[1]{}

\usepackage{graphics,graphicx, color}
\usepackage{dcolumn, amsmath,amssymb}

\begin{document}

\title{Should China build the Great Collider?}

\author{Stephen Hawking\footnote{This memo was completed and distributed in China in early 2018, while Hawking was still active prior to his recent death.}}
\affiliation{DAMTP, CMS, Wilberforce Road, CB3 0WA Cambridge, UK}
\author{Gordon Kane}
\affiliation{Leinweber Center for Theoretical Physics, University of
Michigan, Ann Arbor, MI 48109}

\begin{abstract}
We address the question of the title  from  the scientific,
economic, and cultural points of view, and argue strongly for a
positive answer.  We respond to various issues that have been
raised. \end{abstract}

\maketitle

Throughout the history of human civilization, and especially for the
past four centuries, understanding our physical universe has been a
goal of many people. It is the focus of physics. By the end of the
20th century, we had arrived at a successful, but incomplete,
description of our world: the Standard Models of particle physics
and of cosmology. This description is valid to the highest energies
and to the edges of the universe. It achieves the traditional goals
of physics.

These Standard Models are descriptions, and we do not yet know why
they are correct. In addition, the Standard Models do not include
gravity, particularly a quantum theory of gravity. And they do not
include an explanation of the dark matter of the universe, or why
there were equal amounts of matter and antimatter at the big bang
but today the amount of antimatter in the universe is only one
billionth of the amount of matter, and much more.

The boundaries of physics have changed over the past few decades.
Physicists have become more ambitious. Beginning in the 1970s,
efforts were made to unify the forces into one underlying force
rather than the several we apparently observe. Around the same time,
the idea of supersymmetry was found to be a powerful ingredient in
our potential understanding of such unification. That was reinforced
in the 1980s with the discoveries of inflation and string theory.
Back in the 1920s Ernest Rutherford said ``Don't let me catch anyone
talking about the universe in my department ''. Today it is
different, as Steven Weinberg put it, ``Scientists of the past were
not just like scientists of today who didn't know as much as we do.
They had completely different ideas of what there was to know or how
you go about learning it ''.

Progress in physics can come from new concepts or new tools, such as
new particle colliders or new detectors. Without the CERN Laboratory
Large Hadron Collider (LHC) we would not know about the existence of
the Higgs boson, which changes and sharpens in fundamental ways our
understanding of the universe. For many people it is a source of awe
and comfort to see that humans can understand our universe.

A historical guide as to how not to proceed comes from the U.S.
cancellation of the Superconducting Super Collider (SSC) in 1993.
That has led to the U.S. no longer being the world leader in basic
particle physics, and created an opening for China to move toward
that position. It is well documented that the SSC failed for several
complicated reasons, political and accidental ones, mismanagement,
demanding international participation, and more, with cost overruns
not being a dominant one.

The discovery of the Higgs boson at CERN in 2012 was a wonderful and
major step forward in understanding the universe. It taught us that
the Standard Model of particle physics along with broken symmetries
that allow mass lead to a successful description of our world. The
role of the Higgs interaction is remarkable - if electrons could not
get mass via interacting with the Higgs field then atoms would be
the size of the universe and our world could not exist. Further,
when electrons do get mass via the interaction with the Higgs field,
quantum corrections make them so massive they turn into black holes
unless some new physics yet to be discovered allows them to be
stabilized at their actual mass. The proposed collider will search
for clues to that new physics.

The proposed Chinese collider would have two phases. The first would
be a Circular Electron Positron Collider (CEPC), and the second a
Super Proton Proton Collider (SPPC). Both would be in a long tunnel,
hopefully about 100 km around. The first phase would focus on
learning what the Higgs physics is telling us about a deeper
underlying theory. For example, the LHC data on Higgs boson (h)
decays suggests that the observed Higgs is like a Standard Model
Higgs, even though we know from quantum corrections that the Higgs
cannot actually be a Standard Model one. The several Higgs boson
decay branching ratios are all consistent with being equal to the
Standard Model predictions, even though they could have been very
different. But the LHC data still actually allows quite different
outcomes. The most important decay is $h\to Z+Z$, where Z's are the
bosons that mediate the weak neutral interactions. The ratio of its
LHC value to the prediction is about $1.3\pm 0.3$. The LHC can only
improve that uncertainty a little with further running, while CEPC
could provide an order of magnitude better precision, and really
tell us if the Higgs boson was Standard Model-like or not. The
situation is similar for several other decays. Also, our present
best understanding of the Higgs boson implies that it should be
accompanied by partners. Finding them will require a higher energy
new collider, and searching for them would be a major goal of a
future collider. Better data about Higgs boson properties that could
come from a new collider could lead to truly deeper understanding of
the remarkable role of Higgs physics.

There is an International Linear Collider program in Japan (ILC)
whose goals overlap those of CEPC. There are also studies at CERN
about future colliders. One, CLIC, is a linear electron-positron
collider whose goals would overlap CEPC. In the past there have
often been accelerators or colliders in different countries or
regions with overlapping goals. Scientifically that can be valuable,
and it is surely valuable for all the countries or regions that
construct them, as we discuss below.

One great advantage of CEPC over other proposals, such as the ILC
and CLIC, is that it can have a second phase, called SPPC, to
collide protons at higher energies. The CEPC tunnel will be
available for SPPC, for free. There are strong motivations for
extending the total energy to at least two or three times the LHC
energy, and perhaps ultimately about six or seven times the LHC
energy could be feasible. That would require development of higher
field superconducting magnets. With proton-proton collisions one can
plan for the high luminosity needed to observe signals, and for a
research program lasting decades. One major result to aim for at a
higher energy collider is the data needed to understand how the
Higgs boson itself gets its mass. The second main goal is to search
at significantly higher energies to see what might be discovered.

While no one can be sure what might be discovered eventually at CEPC
or SPPC beyond the guaranteed Higgs physics, one interesting
possibility is the fundamental symmetry called supersymmetry. It
might lead to observable part- ners of the Standard Model particles,
just as the charge conjugation symmetry led to an antiparticle for
every particle. If so, we know their properties are such that they
might be observable at the higher energy SPPC.

Some people have said that the absence of superpartners or other
phenomena at LHC so far makes discovery of superpartners unlikely.
But history suggests otherwise. Once the b quark was found, in 1979,
people argued that ``naturally'' the top quark would only be a few
times heavier. In fact the top quark did exist, but was forty-one
times heavier than the b quark, and was only found nearly twenty
years later. If superpartners were forty-one times heavier than Z
bosons they would be too heavy to detect at LHC and its upgrades,
but could be detected at SPPC. In addition, a supersymmetric theory
has the remarkable property that it can relate physics at our scale,
where colliders take data, with the Planck scale, the natural scale
for a fundamental physics theory, which may help in the efforts to
find a deeper underlying theory. CERN is also studying building a
higher energy proton-proton collider (FCC), with total energy even-
tually about six times that of LHC, perhaps initially only two-three
times LHC. Most likely only one very high energy extension will be
built since it will be fairly costly.

It would be of tremendous benefit to China to build CEPC and its
future upgrades. An essential point to grasp is that when one is at
the frontier of knowledge and understanding, progress requires new
techniques and develop- ments and insights. Otherwise discoveries
would have already been made. Existing techniques and facilities
cannot go further. This has shown up in the past from the LHC in a
number of well documented areas, including inventing the World Wide
Web with its huge impact on economies world-wide and then grid
computing. Someone said imagine that CERN (where the World Wide Web
was invented for particle physics) had one penny for each use, then
particle physics would have all the funding it could use. More
industries include magnet technology and superconducting wire
technology, a multi-billion dollar accelerator industry, a
multi-billion dollar imaging industry that owes its existence to the
development of particle physics detectors, other billion dollar
industries, and many tangible benefits. Such technologies generate
revenues far exceeding the investment for collider construction.

Arguably the third industrial revolution was triggered by the
invention of the World Wide Web at CERN. The requirements for data
acquisition and storage and access, and the materials and
technologies needed for CEPC and SPPC could help lead to the fourth
industrial revolution. For the first decades of the third industrial
revolution High Energy Physics led, and only in recent years
industry has overtaken HEP. History may repeat itself for the
fourth.

About half of all PhD's earned at CERN go to people who move into
indus- tries and areas outside of particle physics, and enrich those
areas. That would happen with CEPC too. A major effect comes because
innovations can lead to start-up companies, but start-ups can be
risky. With LHC to provide an initial market for the products of the
start-ups, they have been far more likely to suc- ceed. That would
be true for a Chinese collider too. New technologies emerge because
particle physics necessarily is at the frontiers, and new approaches
and techniques are needed to interrogate nature more deeply. China
can accelerate the expansion of its economy by investing in a major
collider.

Possibly the largest benefit would be attracting a large number of
bright young Chinese to science and its goals. Those young people
would get excited about many areas of science along the way, and
decide to work in those areas, greatly strengthening the entire
scientific enterprise in China. The Chinese educational system could
handle the challenge of educating many more scientists and benefit
greatly from it.

CEPC may make fundamental new discoveries. Even so, a proton-proton
collider will be needed to discover more or explore properties of
new particles, via a long circular ring with thousands of high field
magnets. Again history provides a guide. The bosons (W and Z and
gluons) that mediate the forces of the Standard Model were
discovered at lower energy facilities. Then CERN built and ran the
LEP electron-positron collider for two decades, studying the
Standard Model and alternatives, and establishing the Standard
Model. Then using the same tunnel, LHC colliding protons at higher
energies was built, and discovered the Higgs boson.

Could there be any alternatives to a higher energy facility to
discover or exclude new particles? People have invented clever
methods to accelerate pro- tons and/or electrons to higher energies,
but unfortunately all approaches have led to luminosities far too
small to discover new physics. At best they lead to a few events per
decade, rather than the tens or hundreds of events a year needed.
Seeing the Higgs boson signal at LHC above backgrounds that could
fake it took over 200,000 events per detector. In the SSC era of the
1980s opponents of the SSC claimed that new magnet technologies
would emerge that would replace the well-established superconducting
magnets, but four decades later such new magnet technologies have
still not arrived, and are unlikely to exist. A description of the
scientific and cultural case for such a collider has been presented
in ``From the Great Wall to the Great Collider: China and the Quest
to Uncover the Inner Workings of the Universe'', by Steve Nadis and
Shing-Tung Yau, published by the International Press of Boston in
2015.

China has several medium size scientific projects, such as the China
Spallation Neutron Source that has just successfully turned on,
operated by the Institute of High Energy Physics and the Institute
of Physics, one of four such facilities in the world. CERN is unique
in high energy physics, the world leader, and a world center for
high energy physics with thousands of physicists from around the
world working at CERN, and large numbers of visitors converging on
CERN to see the laboratory and the detectors. If China built CEPC
and then SPPC as a large science project it could become the
international center of high energy physics, supplanting the role of
CERN. CERN is also studying building such colliders, but only after
a decade or more of upgrading and running the Large Hadron Collider.

The Chinese have so far taken a wise approach to financing a number
of large science facilities, but mostly not at the leading position
in the world, in terms of science, technology, investment scale, and
cultural impact. It is important for China move ahead to take the
leading role, at least in a few selected areas. The CEPC is a good
choice for its scientific importance and technology impact, drawing
on thirty years' experience with the BEPC. Nearly all the costs will
be spent in China. Once China is proceeding, other countries will
join in, stimulating great international collaboration centered
around grand human ambitions, in a spirit of a peace and harmony.

Today collider construction is a mature technology. Cost and time
estimates will be examined by experts, and are likely to be
basically accurate. China's GDP per capita is not yet as high as
that of wealthy nations. But that should not be a reason to back
away from the collider. On the contrary, the collider will provide
work and stimulate economic benefits for many more people. China's
total GDP is now among the largest in the world, and can afford a
future collider. It has been pointed out by Yifang Wang that the
cost of CEPC (and even SPPC) as a fraction of GDP would not exceed
that of the existing and scientifically very successful low energy
Chinese collider, the Beijing Electron Positron Collider (BEPC) when
it was built. Such investments stimulate the technological advances
that raise developing nations to economic leaders. It is important
for China to continue to show wisdom about supporting scientific
research. Funds for a collider should not compete with nor adversely
affect other science funding. Each area should have its funding at a
level that is healthy for its development.

The Chinese particle physics community has matured. It mastered the
low energy collider technology with the Beijing collider, BEPC. Many
Chinese physicists have worked at collider laboratories such as CERN
and Fermilab. If frontier activities are underway in China, foreign
physicists will come to where the action is, and help make any
effort maximally successful. When discoveries come, recognition is
broadly spread. There is some tradition in particle physics for
group leaders and for those whose efforts made the collider
possible, to get Nobel Prizes. For the CERN collider the accelerator
physicist Simon van der Meer and Carlo Rubbia were recipients, and
for the earlier discovery of the charmed quark it was Samuel Ting
and Burton Richter. We can expect Chinese Nobel Prizes.

Could new theoretical concepts or tools emerge that would move
science forward without new collider facilities? Of course new ideas
might lead to new insights. But no matter how elegant a theory might
be, without data we will not know if it really describes or explains
aspects of nature. Without the discovery of the Higgs boson, there
would still be many doubters about the existence of the Higgs field
describing our vacuum state. Results from astrophysics and cosmology
and the cosmic microwave background provide information about
important questions, but no amount of results from these areas could
have told us about the top quark existence or mass, or about the
Higgs physics, or the unification of forces and more. Data will be
crucial to select theories about major issues such as what is the
dark matter, or can we unify and simplify the theory of the forces
and relate the forces to the Higgs mechanism that allows mass, or
what causes the rapid inflation at the beginning of our universe,
and more.

It is remarkable that human cultures could reach the level that
provided data and ideas that have allowed us to take our
understanding of our physical universe to the beginnings of time and
to the edges of the universe. China could take us to the next deeper
level via knowledge obtained from future collider data. The country
that makes the greatest advances in discovering the workings of
nature itself, via the sciences of particle physics and cosmology,
will be permanently remembered in history for glorious achievements.

\section*{Note}
This article, both its original English version and its Chinese
translated version, has been distributed widely in Chinese media
before the death of S.H.. The English and Chinese versions can be
found by the following links:

http://intlpress.sinaapp.com/blog/essay.php?id=16

https://mp.weixin.qq.com/s/87Y89AHg4na4Ai6QQrCNbw

\section*{Acknowledgements}
We are grateful to Nima Arkani-Hamed, Malcolm Perry, Jianming Qian,
Shing-Tung Yau, and Yue Zhao for helpful interactions. The research
of S.H. was supported by the Avery Foundation, and the research of
G.K. was supported in part by Department of Energy grant
de-sc0007859.

\end{document}